\documentclass[aps,prl,twocolumn,groupedaddress,showpacs,showkeys]{revtex4-1}%
\usepackage{amsfonts}
\usepackage{mathtools}
\usepackage{amsmath}
\usepackage{amssymb}
\usepackage{float}
\usepackage{subdepth}
\usepackage{color}
\usepackage{graphicx}
\usepackage{natbib}
\usepackage{hyperref}
\usepackage{url}
\usepackage{soul}%
\providecommand{\U}[1]{\protect\rule{.1in}{.1in}}
\hypersetup{colorlinks, linkcolor={blue}, citecolor={blue}, urlcolor={blue}}
\begin{document}

\title{Nonlocal Drag Thermoelectricity Generated by Ferroelectric Heterostructures}
\author{Ping Tang$^{1}$}
\author{Ken-ichi Uchida$^{2,3}$}
\author{Gerrit E. W. Bauer$^{1,3,4,5,6}$}
\affiliation{$^1$WPI-AIMR, Tohoku
University, 2-1-1 Katahira, Sendai 980-8577, Japan}
\affiliation{$^2$National Institute for Materials Science, Tsukuba 305-0047, Japan}
\affiliation{$^3$Institute for Materials Research, Tohoku University,
2-1-1 Katahira, Sendai 980-8577, Japan} \affiliation{$^4$Center for
Spintronics Research Network, Tohoku University, Sendai 980-8577, Japan}
\affiliation{$^5$Zernike Institute for Advanced Materials, University of
Groningen, 9747 AG Groningen, Netherlands}
\affiliation{$^6$Kavli Institute for Theoretical Sciences, University of the Chinese Academy of Sciences,
Beijing 10090, China}
\begin{abstract}
The \textquotedblleft ferron\textquotedblright\ excitations of the
electric-dipolar order carry energy as well as electric dipoles. Here we
predict a nonlocal ferron drag effect in a ferroelectric on top of a metallic
film: An electric current in the conductor generates a heat current in the
ferroelectric by long-range charge-dipole interactions. The non-local Peltier
and its reciprocal Seebeck effect can be controlled by electric gates and
detected thermographically. We predict large effects for van der Waals
ferroelectric films on graphene.

\end{abstract}
\maketitle

%%%%%%%%%%%%%%%%%%%%%%%%%%%%
%%%%%%%%%%%%%%%%%%%%%%%%%%%%
The electron-electron interaction between closely spaced two-dimensional
electron gases (2DEGs) gives rise to \textit{non-local} Coulomb drag effects
\cite{gramila1991mutual,jauho1993coulomb,narozhny2016coulomb}, in which a
current in an active layer induces a voltage over the passive one. The concept
of Coulomb drag has been extended to other systems and interactions. A
\textit{local} drag effect by the electron-phonon interaction contributes to
the thermopower in bulk conductors
\cite{bailyn1967phonon,cantrell1987calculation,lyo1988low,vavro2003thermoelectric}
and also a non-local drag effect can be mediated by phonons in the spacer
between the 2DEGs \cite{tso1992direct,bonsager1998frictional,noh1999phonon}.
In ferromagnetic metals, magnons, the quasiparticle excitations of local
magnetization, transfer their momenta to conduction electrons by the exchange
interaction. This local magnon drag effect enhances the Seebeck and Peltier
coefficients
\cite{bailyn1962maximum,blatt1967magnon,PhysRevB.13.2072,costache2012magnon,flebus2016landau,watzman2016magnon}%
. The voltage in one layer induced by a current in the other in a heavy
metal/ferromagnetic insulator/heavy metal stack is a non-local drag effect
caused by spin Hall effect
\cite{zhang2012magnon,wu2016observation,li2016observation}. Theory predicts
that magnons in magnetic films separated by a vacuum barrier experience a
non-local drag effect by the magnetodipolar interaction \cite{liu2016nonlocal}%
. The magnetodipolar interaction can also mediate an energy transfer through
an air gap \cite{Kainuma2021}, but a non-local magnon drag effect has not yet
been observed.

Ferroelectrics exhibit an electrically switchable spontaneous polarization
that orders below a Curie temperature. Recently, we introduced
\textquotedblleft ferrons\textquotedblright, the bosonic excitations of
ferroelectric order that carry elementary electric dipoles in the presence of
transverse \cite{Bauer2021,Tang2022} or longitudinal fluctuations
\cite{arXiv:2203.06367}. A direct experimental observation of the predicted
polarization and heat transport phenomena, e.g. by the transient Peltier
effect \cite{Bauer2021} and associated stray fields \cite{Tang2022}, may not
be so simple, however.

\begin{figure}[ptb]
\centering
\par
\includegraphics[width=8.2cm]{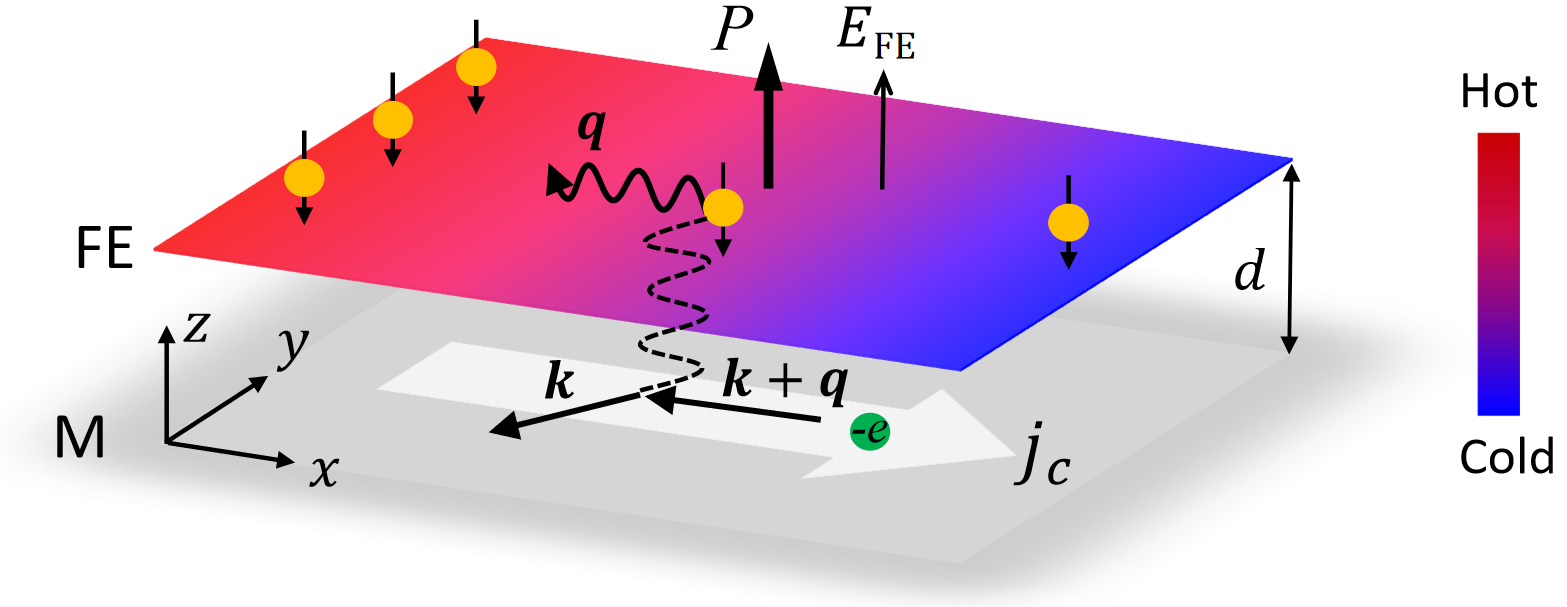}\newline\caption{A schematic of the
nonlocal ferron-drag Peltier effect between an extended metallic (M) and a
perpendicularly polarized and electrically insulating ferroelectric (FE) film.
A charge current ($j_{c}$) in the active M sheet transfers its linear momentum
to the ferrons in the FE by the electrostatic interaction, leading to heat
accumulations at the FE edges. The orange balls represent the ferrons, while
the small black arrows are the ferron dipoles that oppose the ferroelectric
order and can couple with an out-of-plane electric field ($E_{\mathrm{FE}}$).}%
\label{Fig-1}%
\end{figure}

Here we pursue ideas to simplify the detection of ferronic effects via
non-local thermoelectric drag effects in bilayers of a ferroelectric and a
metal, which opens new strategies for heat-to-electricity conversion. We
consider a film of a perpendicularly polarized ferroelectric insulator on top
of an extended metallic sheet that experiences a \textquotedblleft ferron
drag\textquotedblright\ in the form of a non-local Peltier effect, i.e., a
heat current in the ferroelectric generated by an electric current in the
metal film (see figure~\ref{Fig-1}). We assume that the electric dipoles are
all located in a common plane and that the electrons in the metal move in a
parallel plane. This two-dimensional (2D) assumption is valid when the two
films are separated by a distance $d$ much larger than their thickness, but
certainly appropriate when the conductor is, e.g., graphene and the
ferroelectric a van der Waals mono- or bilayer \cite{Chang2016,Liu2016,
Li2017, Yang2018, Fei2018, Yuan2019, Yasuda2021, Wang2022}.

The linear response relations of transport or Ohm's Law in our bilayer (in the
$x$-direction) connect four driving forces, i.e. an in-plane electric field
$E_{\mathrm{M}}$ in the metal, a gradient of an out-of-plane electric field
$\partial E_{\mathrm{FE}}$ in the ferroelectric, and independent temperature
gradients in the two films, with the charge current $j_{c}$ in the metal,
polarization current $j_{p}$ in the ferroelectric and the heat currents
$j_{q}^{(\mathrm{M})}$ and $j_{q}^{(\mathrm{FE})}$:
\begin{equation}
\left(
\begin{array}
[c]{c}%
-j_{c}\\
j_{q}^{(\mathrm{M})}\\
-j_{p}\\
j_{q}^{(\mathrm{FE})}%
\end{array}
\right)  =\left(
\begin{array}
[c]{cccc}%
L_{11} & L_{12} & L_{13} & L_{14}\\
L_{12} & L_{22} & L_{23} & L_{24}\\
L_{13} & L_{23} & L_{33} & L_{34}\\
L_{14} & L_{24} & L_{34} & L_{44}%
\end{array}
\right)  \left(
\begin{array}
[c]{c}%
-E_{\mathrm{M}}\\
-\partial\ln T_{\mathrm{M}}\\
\partial E_{\mathrm{FE}}\\
-\partial\ln T_{\mathrm{FE}}%
\end{array}
\right)
\end{equation}
where we already inserted the Onsager-Kelvin relation $L_{ij}=L_{ji}$ between
the off-diagonal transport coefficients. We focus here on the steady state
with finite $E_{\mathrm{M}}$ that induces polarization and heat currents in
the ferroelectric. In the following, we disregard small thermoelectric effects
in the metal, thermal leakage between the films, and electric field gradients
$\partial E_{\mathrm{FE}}$ at the edges of the ferroelectric. The task then
reduces to the calculation of the polarization drag $\vartheta_{D}\equiv
L_{13}/L_{11}$ as well as the thermoelectric effects summarized by
\begin{equation}
\left(
\begin{array}
[c]{c}%
-j_{c}\\
j_{q}^{(\mathrm{FE})}%
\end{array}
\right)  =\left(
\begin{array}
[c]{cc}%
L_{11} & L_{14}\\
L_{14} & L_{44}%
\end{array}
\right)  \left(
\begin{array}
[c]{c}%
-E_{\mathrm{M}}\\
-\partial\ln T_{\mathrm{FE}}%
\end{array}
\right)
\end{equation}
in which we identify the non-local Peltier coefficient $\pi_{D} =
L_{14}/L_{11}$ and thermopower $s_{D}=\pi_{D}/T_{\text{FE}}$. The electrical
conductivity $\sigma=L_{11}$ is also affected by the equilibrium fluctuations
of the nearby ferroelectric.

The conduction electrons in the metallic layer interact with the electric
polarization $\mathbf{P}\left(  \mathbf{r}\right)  =P\left(  \mathbf{r}%
_{\Vert}\right)  \delta\left(  z-d\right)  \mathbf{\hat{z}}$ of the
ferroelectric at $z=d$ by the electrostatic energy
\begin{equation}
\mathcal{H}_{\mathrm{int}}=-\int\mathbf{E}_{\mathrm{el}}\left(  \mathbf{r}%
\right)  \cdot\mathbf{P}(\mathbf{r})d\mathbf{r,}%
\end{equation}
where
\begin{equation}
\mathbf{E}_{\mathrm{el}}(\mathbf{r})=-\frac{e}{4\pi\epsilon_{r}\epsilon_{0}%
}\int d\mathbf{r}^{\prime}\frac{n(\mathbf{r}^{\prime})}{|\mathbf{r}%
-\mathbf{r}^{\prime}|^{3}}(\mathbf{r}-\mathbf{r}^{\prime}) \label{Ele}%
\end{equation}
is the Hartree field of the electrons, $-e$ the electron charge,
$n(\mathbf{r})=n(\mathbf{r}_{\Vert})\delta\left(  z\right)  $ the electron
density in the metal at $z=0$ and $\epsilon_{r}$ the relative permittivity of
the separating barrier. Substituting Eq.~(\ref{Ele}) leads to
\begin{equation}
\mathcal{H}_{\mathrm{int}}=\frac{ed}{4\pi\epsilon_{r}\epsilon_{0}}\int\int
d\mathbf{r}_{\Vert}d\mathbf{r}_{\Vert}^{\prime}\frac{P(\mathbf{r}_{\Vert
})n(\mathbf{r}_{\Vert}^{\prime})}{[(\mathbf{r}_{\Vert}-\mathbf{r}_{\Vert
}^{\prime})^{2}+d^{2}]^{3/2}}. \label{Inter}%
\end{equation}
where $P(\mathbf{r}_{\Vert})$ and $n(\mathbf{r}_{\Vert})$ represent the 2D
polarization and electron density in units of C/m and m$^{-2}$, respectively.

We model the ferroelectric by the Landau-Ginzburg-Devonshire free energy
\cite{Devonshire1949,Devonshire1951}
\begin{equation}
F=\left(  \frac{g}{2}(\boldsymbol{\nabla}P)^{2}+\frac{\alpha}{2}P^{2}%
+\frac{\beta}{4}P^{4}+\frac{\lambda}{6}P^{6}-E_{\mathrm{FE}}P\right)
\label{free}%
\end{equation}
where $\alpha=\alpha_{0}(T-T_{c})$, $\beta$ and $\lambda>0$ are the Landau
coefficients, $T_{c}$ the Curie-Weiss temperature, $g>0$ the Ginzburg parameter
accounting for the energy cost of an inhomogeneous polarization, and
$E_{\text{FE}}$ is an out-of-plane electric field acting on the ferroelectric
order. The phase transition for $\beta<0$ ($\beta>0$) is first (second)-order.
A uniform spontaneous polarization $P_{0}$ minimizes $F$ by $\alpha
P_{0}+\beta P_{0}^{3}+\lambda P_{0}^{5}=E_{\mathrm{FE}}$, which gives
$P_{0}^{2}=[ -\beta+(\beta^{2}-4\alpha\lambda)^{1/2}]/(2\lambda)$ when
$E_{\mathrm{FE}}=0$. The non-linear static dielectric susceptibility with the
field reads
\begin{equation}
\chi\left(  E_{\mathrm{FE}}\right)  =\frac{\partial P_{0}\left(
E_{\mathrm{FE}}\right)  }{\partial E_{\mathrm{FE}}}=\frac{1}{\alpha+3\beta
P_{0}^{2}\left(  E_{\mathrm{FE}}\right)  +5\lambda P_{0}^{4}\left(
E_{\mathrm{FE}}\right)  }.
\end{equation}
Small fluctuations $\delta P(\mathbf{r}_{\Vert},t)=P(\mathbf{r}_{\Vert
},t)-P_{0}$ can be quantized as \cite{arXiv:2203.06367}
\begin{equation}
\delta P(\mathbf{r}_{\Vert},t)=\sqrt{\frac{\hbar}{2m_{p}A}}\sum_{\mathbf{q}%
}\frac{\hat{a}_{\mathbf{q}}e^{i\mathbf{q}\cdot\mathbf{r}_{\Vert}}}%
{\sqrt{\omega_{q}}}+\mathrm{H.c.}, \label{pflu}%
\end{equation}
where $m_{p}$ is the polarization inertia that depends on the ionic masses
$M_{i}$ and Born effective charges $Q_{i}$ in the unit cell of area $A_{0}$ as
$m_{p}=A_{0}(\sum_{i}Q_{i}^2/M_{i})^{-1}$ \cite{Sivasubramanian2004}, $A$ the
area of the ferroelectric sheet (assumed to be the same as the metal) and
$\hat{a}_{\mathbf{q}}$ ($\hat{a}_{\mathbf{q}}^{\dagger}$) the annihilation
(creation) operator of ferrons with the dispersion relation
\begin{equation}
\omega_{q}=\left(  \frac{gq^{2}+\chi\left(  E_{\mathrm{FE}}\right)  ^{-1}%
}{m_{p}}\right)  ^{1/2}.
\end{equation}
The electric dipole carried by a single ferron is then identified as
\cite{arXiv:2203.06367}
\begin{equation}
\delta p_{q}=-\frac{\partial\hbar\omega_{q}}{\partial E_{\mathrm{FE}}}%
=\frac{\hbar}{2m_{p}\omega_{q}}\frac{\partial\ln\chi}{\partial P_{0}}<0
\end{equation}
where the negative sign indicates its opposite direction to the ferroelectric order.

In 2D momentum space Eq.~(\ref{Inter}) now reads
\begin{equation}
\mathcal{H}_{\mathrm{int}}=\frac{e}{2\epsilon_{r}\epsilon_{0}A}\sum
_{\mathbf{q}}e^{-dq}n(\mathbf{q})\int d\mathbf{r}_{\Vert}\delta{P}%
(\mathbf{r}_{\Vert})e^{i\mathbf{q}\cdot\mathbf{r}_{\Vert}}%
\end{equation}
where we dropped a constant energy shift related to $P_{0}$ and $n(\mathbf{q}%
)=\sum_{\mathbf{k}\nu\nu^{\prime}}F_{\mathbf{k}\nu}^{\dagger}F_{(\mathbf{k}%
+\mathbf{q})\nu^{\prime}}\hat{c}_{\mathbf{k}\nu}^{\dagger}\hat{c}%
_{(\mathbf{k}+\mathbf{q})\nu^{\prime}}$ is the Fourier component of the 2D
electron density in terms of the field operators $\hat{c}_{\mathbf{k}\nu
}^{\dagger}$ and $\hat{c}_{\mathbf{k}\nu}$ with momentum $\mathbf{k}$, band
index $\nu$ and the corresponding spinor wave functions $F_{\mathbf{k}\nu}$.
$F_{\mathbf{k}^{\prime}\nu^{\prime}}^{\dagger}F_{\mathbf{k}\nu}=(e^{i(\theta
_{\mathbf{k}^{\prime}}-\theta_{\mathbf{k}})}+\nu\nu^{\prime})/2$
($=\delta_{\nu,\nu^{\prime}}$) for graphene (normal metals), where $\tan
\theta_{\mathbf{k}}=k_{y}/k_{x}$ and $\nu=+1$ and $\nu=-1$ indicate the
conduction and valence bands, respectively \cite{Ando2006,Hwang2009}.
Substituting Eq.~(\ref{pflu}) yields
\begin{equation}
\mathcal{H}_{\mathrm{int}}=\sum_{\mathbf{k}\mathbf{q}\nu\nu^{\prime}}
V_{\mathbf{k}\mathbf{q}}(\nu^{\prime},\nu)\hat{c}_{(\mathbf{k}+\mathbf{q})\nu^{\prime}}^{\dagger
}\hat{c}_{\mathbf{k}\nu}\hat{a}_{\mathbf{q}}+\mathrm{H.c.},
\end{equation}
where
\begin{equation}
V_{\mathbf{k}\mathbf{q}}(\nu^{\prime},\nu)=\frac{e}{2\epsilon_{r}\epsilon_{0}}\sqrt{\frac{\hbar
}{2m_{p}A}}\frac{e^{-dq}}{\sqrt{\omega_{q}}}F_{(\mathbf{k}+\mathbf{q}%
)\nu^{\prime}}^{\dagger}F_{\mathbf{k}\nu}%
\end{equation}
is the bare inelastic scattering amplitude of the electrons. 

The screening by the conduction electrons and electric dipoles is a many-body
problem in which $V_{\mathbf{k}\mathbf{q}}(\nu^{\prime},\nu)\rightarrow V_{\mathbf{k}\mathbf{q}}(\nu^{\prime},\nu)/\epsilon\left(  q,\omega\right)  $
and $\epsilon(q,\omega)$ is the dielectric function. At sufficiently high
conduction electron densities, the ferron energies are small compared to the
Fermi energy and we may adopt static screening $\omega\rightarrow0$. For $q
<1/(2d)\lesssim2k_{F}$, where $k_{F}$ is the Fermi wave vector, it is
sufficient to adopt the Thomas-Fermi screening approximation \cite{Stern1967,
Ando1982, Ando2006, Hwang2007, Hwang2009, Sarma2011}, i.e.,
\begin{equation}
V_{\mathbf{k}\mathbf{q}}(\nu^{\prime},\nu)\rightarrow U_{\mathbf{k}\mathbf{q}}(\nu^{\prime}\nu)=\frac{V_{\mathbf{k}\mathbf{q}}(\nu^{\prime}\nu)}{1+q_{\mathrm{TF}}/q}%
\end{equation}
where $q_{\mathrm{TF}}=e^{2}D_{F}/(2\epsilon_{r}\epsilon_{0})$ is the 2D
Thomas-Fermi wave vector in terms of the density of state $D_{F}$ at Fermi
level. The screening by the ferroelectric dipoles is negligibly small compared
to that of the free electrons when the ferroelectric sheet is sufficiently
thin. The screening then does not depend on $d$.

We consider now the effect of a charge current $j_{c}$ driven by an electric
filed ($E_{\mathrm{M}}$) along the $x$ direction in the metallic sheet that
deforms the electron distribution function $f_{\mathbf{k\nu}}$ from the
Fermi-Dirac form $f_{\mathbf{k}\nu}^{(0)}=[\exp\left(  (\varepsilon
_{\mathbf{k}\nu}-\varepsilon_{F})/k_{B}T\right)  +1]^{-1}$ in momentum space,
where $\varepsilon_{\mathbf{k\nu}}$ is the electronic band structure,
$\varepsilon_{F}$ the Fermi energy, $T$ the temperature, and $k_{B}$
Boltzmann's constant. Within relaxation time approximation the linearized
Boltzmann equation in the metal reads
\begin{equation}
f_{\mathbf{k\nu}}=f_{\mathbf{k}\nu}^{(0)}+e\tau_{e}v_{\mathbf{k}\nu}^{(x)}
E_{M} \frac{\partial f_{\mathbf{k}\nu}^{(0)}}{\partial\varepsilon
_{\mathbf{k}\nu}} \label{fk}%
\end{equation}
where $\tau_{e}$ is the relaxation time and $v_{\mathbf{k}\nu}^{(x)}%
=\partial\varepsilon_{\mathbf{k}\nu}/\partial\hbar k_{x}$ the group velocities
in transport $(x)$ direction, with $v_{\mathbf{k}\nu}^{(x)}\rightarrow\hbar
k_{x}/m_{e}$ $\left(  \nu v_{F}k_{x}/\vert\mathbf{k}\vert\right)  $ for a free
electron gas with effective mass $m_{e}$ (or a Dirac cone of graphene with
Fermi velocity $v_{F}$). The associated electric current density reads
$j_{c}=\sigma E_{M}$, where
\begin{equation}
\sigma=\frac{e^{2}\tau_{e} \delta}{A}\sum_{\mathbf{k}}(v_{\mathbf{k}\nu}%
^{(x)})^{2}\left(  - \frac{\partial f_{\mathbf{k}\nu}}{\partial\varepsilon
_{\mathbf{k}\nu}}\right)
\end{equation}
is the electrical conductivity and $\delta$ includes the spin and valley degeneracies.

In the Supplemental Material A \cite{SM} we derive a ferron-electron
scattering contribution that drastically reduces the $\tau_{e}$ at the Curie
temperature of the ferroelectric. The observation of the predicted critical
enhancement of the scattering rate would provide a proof of ferron excitations
independent of the thermoelectric effects discussed in the following.

The bosonic ferron distribution function $N_{\mathbf{q}}$ in the ferroelectric
is governed by another linearized Boltzmann equation
\cite{arXiv:2203.06367,Bauer2022}. Far from the edges and in the absence of
temperature or effective field gradients, the steady state distribution reads
\begin{equation}
N_{\mathbf{q}}=N_{q}^{(0)}+\tau_{f}\left.  \frac{\partial N_{\mathbf{q}}%
}{\partial t}\right\vert _{\mathrm{int}} \label{Bol}%
\end{equation}
where $N_{q}^{(0)}=\left[  \exp\left(  \hbar\omega_{q}/k_{B}T\right)
-1\right]  ^{-1}$ is the equilibrium Planck distribution, $\tau_{f}$ the
ferron relaxation time. The new ingredient is the collision integral $\left.
\partial N_{\mathbf{q}}/\partial t\right\vert _{\mathrm{int}}$, which by the
current in the metal and via the interlayer interaction $U_{\mathbf{q}}$
renders $N_{\mathbf{q}}\neq N_{-\mathbf{q}}.$ The electrons scatter from
occupied to empty states$\mathbf{,}$ creating and annihilating a ferron in the
process. According to Fermi's Golden Rule
\begin{align}
\left.  \frac{\partial N_{\mathbf{q}}}{\partial t}\right\vert _{\mathrm{int}}
&  =\frac{2\pi\delta}{\hbar}\sum_{\mathbf{k}}\left\vert U_{\mathbf{k}\mathbf{q}}\right\vert
^{2}\left[  (1+N_{\mathbf{q}})f_{\left(  \mathbf{k}+\mathbf{q}\right)  \nu
}(1-f_{\mathbf{k}\nu})\right. \nonumber\\
&  \left.  -N_{\mathbf{q}}f_{\mathbf{k}\nu}(1-f_{\left(  \mathbf{k}%
+\mathbf{q}\right)  \nu})\right]  \delta(\varepsilon_{\mathbf{k}\nu
}-\varepsilon_{\left(  \mathbf{k}+\mathbf{q}\right)  \nu}+\hbar\omega_{q})
\label{FGR}%
\end{align}
while energy and momentum are conserved. Here insignificant interband
processes ($\nu\neq\nu^{\prime}$) have been discarded. To leading order, we
may replace $N_{\mathbf{q}}$ on the r.h.s. of Eq.~(\ref{FGR}) with
$N_{q}^{(0)}$ and substitute the distribution function of the field-biased
conductor Eq.~(\ref{fk}):
\begin{align}
N_{\mathbf{q}}  &  =N_{q}^{(0)}+\frac{2\pi\delta\tau_{f}}{\hbar}\frac{\partial
N_{q}^{(0)}}{\partial\hbar\omega_{q}}%
\sum_{\mathbf{k}}\left\vert U_{\mathbf{k}\mathbf{q}}\right\vert ^{2}(f_{\left(  \mathbf{k}+\mathbf{q}\right)  \nu}^{(0)}%
-f_{\mathbf{k}\nu}^{(0)})\nonumber\\
&  \times e\tau_{e}E_{M} (v_{\mathbf{k}+\mathbf{q}}^{(x)}-v_{\mathbf{k}}%
^{(x)})\delta(\varepsilon_{\mathbf{k}\nu}-\varepsilon_{\left(  \mathbf{k}%
+\mathbf{q}\right)  \nu}+\hbar\omega_{q}).
\end{align}

We can now derive the non-local Peltier $\pi_{D}=-j_{q}^{(\mathrm{FE})}/j_{c}$
and polarization drag $\vartheta_{D}=j_{p}/j_{c}$ coefficients by evaluating
the heat and polarization currents for the deformed ferron distribution
functions by $j_{q}^{(\mathrm{FE})}=A^{-1}\sum_{\mathbf{q}}u_{\mathbf{q}%
}^{(x)}N_{\mathbf{q}}\hbar\omega_{q}$ and $j_{p}=A^{-1}\sum_{\mathbf{q}%
}u_{\mathbf{q}}^{(x)}N_{\mathbf{q}}\delta p_{q}$, respectively, where
$u_{\mathbf{q}}^{(x)}=\partial\omega_{q}/\partial q_{x}$ is the ferron group
velocity along $x$ direction:
\begin{align}
\pi_{D}= &  \frac{2\pi e\tau_{f}\tau_{e}\delta}{\sigma\hbar A}\sum
_{\mathbf{k}\mathbf{q}}\hbar\omega_{q}u_{\mathbf{q}}^{\left(  x\right)
}(v_{\mathbf{k}+\mathbf{q}}^{(x)}-v_{\mathbf{k}}^{\left(  x\right)  }%
)\frac{\partial N_{q}^{(0)}}{\partial\hbar\omega_{q}}\left\vert U_{\mathbf{k}\mathbf{q}}
\right\vert ^{2}\nonumber\\
&  \times(f_{\mathbf{k}+\mathbf{q}\nu}^{(0)}-f_{\mathbf{k}\nu}^{(0)}%
)\delta(\varepsilon_{\mathbf{k}\nu}-\varepsilon_{\left(  \mathbf{k}%
+\mathbf{q}\right)  \nu}+\hbar\omega_{q})\\
\vartheta_{D}= &  \frac{2\pi e\tau_{f}\tau_{e}\delta}{\sigma\hbar A}%
\sum_{\mathbf{k}\mathbf{q}}\delta p_{q}u_{\mathbf{q}}^{\left(  x\right)
}(v_{\mathbf{k}+\mathbf{q}}^{(x)}-v_{\mathbf{k}}^{\left(  x\right)  }%
)\frac{\partial N_{q}^{(0)}}{\partial\hbar\omega_{q}}\left\vert U_{\mathbf{k}\mathbf{q}}%
\right\vert ^{2}\nonumber\\
&  \times(f_{\mathbf{k}+\mathbf{q}\nu}^{(0)}-f_{\mathbf{k}\nu}^{(0)}%
)\delta(\varepsilon_{\mathbf{k}\nu}-\varepsilon_{\left(  \mathbf{k}%
+\mathbf{q}\right)  \nu}+\hbar\omega_{q}).
\end{align}

We proceed by adopting the quasi-elastic approximation, i.e., $\delta
(\varepsilon_{\mathbf{k}\nu}-\varepsilon_{\left(  \mathbf{k}+\mathbf{q}%
\right)  \nu}+\hbar\omega_{q})\approx\delta(\varepsilon_{\mathbf{k}\nu
}-\varepsilon_{\left(  \mathbf{k}+\mathbf{q}\right)  \nu})$, assuming that the
Fermi energy is much larger than that of the ferrons ($\lesssim10\,$meV)
\cite{arXiv:2203.06367}. This is the case in graphene with homogeneous
electron densities $n_{0}>10^{12}\,$cm$^{-2}$ and Fermi energies
$\varepsilon_{F}>0.11\,$eV \cite{Sarma2011}. At $k_{B}T\ll\varepsilon_{F}$,
$f_{\mathbf{k}\nu}-f_{(\mathbf{k}+\mathbf{q})\nu}\simeq\hbar\omega_{q}%
\delta(\varepsilon_{\mathbf{k}\nu}-\varepsilon_{F})$ and we find
\begin{align}
\pi_{D}\simeq &  \frac{e\tau_{f}g\hbar^{3}D_{F}^{2}}{32\delta m_{p}%
^{2}\epsilon_{0}^{2}n_{0}k_{F}k_{B}T}\int_{0}^{2k_{F}}\frac{\cos^{2}(\theta/2)
q^{2}dq}{\sqrt{1-(q/2k_{F})^{2}}}\nonumber\\
&  \times\frac{e^{-2dq}}{(1+q_{\mathrm{TF}}/q)^{2}}\mathrm{csch}^{2}\left(
\frac{\hbar\omega_{q}}{2k_{B}T}\right)  . \label{nonp}%
\end{align}
In contrast to the free electron gas there is a factor $\cos^{2}(\theta/2)$
that arises from the overlap $\vert F_{(\mathbf{k}+\mathbf{q})\nu}^{\dagger
}F_{\mathbf{k}\nu}\vert^{2}$, where $\theta$ is the scattering angle
determined by $q=2k_{F}\sin(\theta/2)$. A similar expression can be derived
for $\vartheta_{D}$ by replacing $\hbar\omega_{q}$ with $\delta p_{q}$.

The spatial separation limits the momentum transfer exponentially via the
factor $\exp(-2dq)$ to $q <1/(2d)$. At large distances with $k_{F}d\gg1$,
$q_{\mathrm{TF}}d\gg1$ and $d\gg l\equiv\sqrt{g\chi}$, only the ferrons
located near the center of Brillouin zone contribute and
\begin{align}
\pi_{D}  &  \approx\frac{3\pi}{8\delta^{2}}\frac{\tau_{f}\omega_{0}}%
{(k_{F}d)^{3}}\left(  \frac{l}{d}\right)  ^{2}\left(  \frac{\hbar^{2}}%
{e^{3}m_{p}}\right)  \xi_{0}\mathrm{csch}^{2}\left(  \frac{\xi_{0}}{2}\right)
\nonumber\\
&  \simeq\frac{3\pi}{2\delta^{2}}\frac{\tau_{f}\omega_{0}}{(k_{F}d)^{3}%
}\left(  \frac{l}{d}\right)  ^{2}\left(  \frac{\hbar^{2}}{e^{3}m_{p}}\right)
\left\{
\begin{array}
[c]{c}%
\xi_{0}e^{-\xi_{0}},\\
\xi_{0}^{-1},
\end{array}
\text{ for }%
\begin{array}
[c]{c}%
\xi_{0}\gg1\\
\xi_{0}\ll1
\end{array}
\right. \nonumber\\
\vartheta_{D}  &  \approx\frac{\pi_{D}}{2}\frac{\partial\chi}{\partial P_{0}}
\label{DPel}%
\end{align}
where $\xi_{0}=\hbar\omega_{0}/k_{B}T$ and $\omega_{0}=(\chi m_{p})^{-1/2}$
the ferron gap. $l\equiv\sqrt{g\chi}$ is the coherence length of the
ferroelectric order, a measure of the ferroelectric domain wall width
\cite{Ishibashi1989,Ishibashi1990}. Since magnetic domain wall widths that
scale like $\sim\sqrt{J/K},$ where $J$ is the exchange interaction and $K$ is
the anisotropy that governs the magnon gap, $\chi^{-1}$ plays the role of the
anisotropy by stiffening the ferroelectric order.

The $T/d^{5}$ scaling relation at large distances and elevated temperatures
for the drag efficiency differs from that of the Coulomb drag effect between
two metallic sheets ($\sim T^{2}/d^{4}$)
\cite{gramila1991mutual,jauho1993coulomb}. We can trace the difference to the
faster decay of electron-dipole interactions $(\sim r^{-2})$ compared to those
between charges $(\sim r^{-1})$ as a function of distance while the Planck
distribution of the ferrons compared to the Fermi distribution of electrons
leads to an increased phase space for scatterings at low temperatures
($k_{B}T\ll\varepsilon_{F}$).

\begin{figure}[ptb]
\centering
\par
\includegraphics[width=7.6cm]{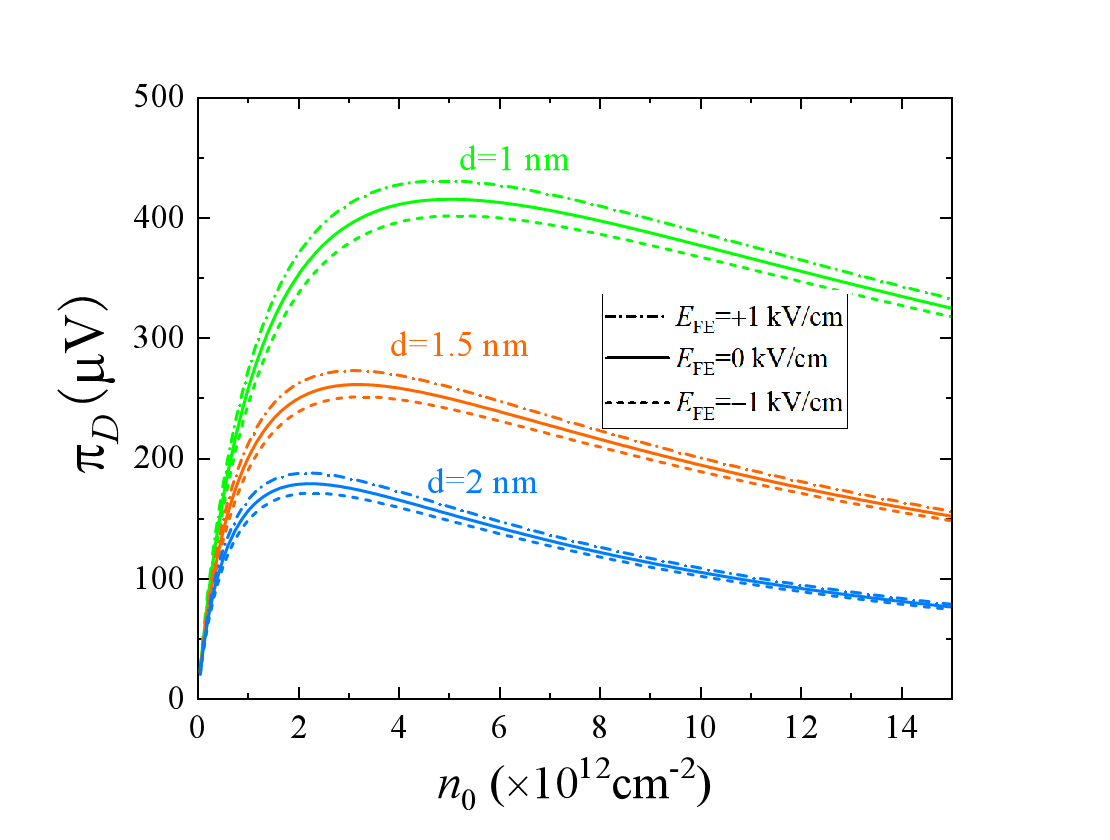}\newline\caption{The nonlocal
ferron-drag Peltier coefficient ($\pi_{D}$) as a function of the electron
concentration ($n_{0}$) in graphene for various interlayer distance ($d$). The
$\pi_{D}$ exhibit an asymmetric dependence on the relative direction of an
external field ($E_{\text{FE}}$) to the ferroelectric order.}%
\label{Fig-2}%
\end{figure}

For a numerical estimate, we consider here a bilayer composed of graphene and
van der Waals ferroelectric monolayer and separated by an inert h-BN layer
with (out-of-plane) $\epsilon_{r}=3.76$ \cite{Laturia2018}. In graphene
$\varepsilon_{\mathbf{k}\nu}=\nu\hbar v_{F}|\mathbf{k}|$ with $v_{F}=10^{8}%
$\thinspace cm/s, $D_{F}=2\varepsilon_{F}/(\pi\hbar^{2} v_{F}^{2})$,
$\delta=4$ and $k_{F}=(4\pi n_{0}/\delta)^{1/2}$ \cite{Sarma2011}. The
parameters for the ferroelectric are adopted as: $\tau_{f}=1\,$ps,
$T_{c}=326\,$K, $\alpha_{0}=1.54\times10^{3}\,$VK$^{-1}$/pC, $\beta
=1.48\times10^{5}\,$Vcm$^{2}$/pC$^{3}$, $\lambda=2.75\times10^{4}\,$Vcm$^{4}%
$/pC$^{5}$ and $g=0.33\,$Vm$^{2}$/C, and $m_{p}=10^{-8}\,$Vs$^{2}$/C, which
are close to those of the monolayer SnSe with in-plane polarization
\cite{Fei2016}. Figure~\ref{Fig-2} shows the ferron-drag Peltier coefficient
as a function of graphene excess electron density ($n_{0}$) for various
interlayer distances $d$ at room temperature. $\pi_{D}$ has a maximum at an
optimal $n_{0}$ that decreases with $d$ because a larger $n_{0}$ increases the
electron-ferron scattering for small $n_{0}$ while the increased screening
wins at larger densities, which is easier for larger $d$. $\pi_{D}$ depends
not only on the strength, but also on the direction of an external electric
field (below the coercive field), i.e., $\pi_{D}$ is reduced (enhanced) by the
positive (negative) field along the ferroelectric order, because of the fact
that the ferrons carry nonzero electric dipoles.

The drag effect results in heat and polarization accumulations in the
ferroelectric (see figure~\ref{Fig-1}). Assuming that both films are thermally
isolated, a temperature gradient $T_{\mathrm{FE}}(x)=T_{0}+\partial
T_{\mathrm{FE}}\left(  x-L/2\right)  $ emerges in a ferroelectric with length
$L$, where $T_{0}$ is the ambient temperature. The open circuit condition for
the heat current, i.e., $j_{q}^{(\mathrm{FE})}=-\pi_{D}j_{c}-\kappa
_{\mathrm{FE}}\partial T_{\mathrm{FE}}=0$, leads to $\partial T_{\mathrm{FE}%
}=\left(  -\pi_{D}/\kappa_{\mathrm{FE}}\right)  j_{c}$, where $\kappa
_{\mathrm{FE}}$ is the 2D thermal conductivity of the ferroelectric sheet (in
units of W/K). The polarization accumulation $\Delta P(x)$ vanishes except for
the neighborhood of the edges on the scale of the polarization relaxation
length \cite{Bauer2021}.

With $d=1\,$nm, $n_{0}=10^{13}\,$cm$^{-2}$, we have $\pi_{D}=367$%
\thinspace$\mathrm{\mu V}$ at the room temperature. The maximum current
density in graphene is limited by self-heating to $\sim30\,$A/cm
\cite{Liao2011}, but even for $j_{c}=3.4\times10^{-2}\,$A/cm (or a bulk
current density $j_{c}^{(b)}=10^{6}\,$A/cm$^{2}$) this modest Peltier
coefficient generates a large temperature gradient $\partial T_{\mathrm{FE}%
}=5$\thinspace K/$\mathrm{\mu m}$ for $\kappa_{\mathrm{FE}}=2.5\times
10^{-10}\,$W/K (or bulk $\kappa_{\mathrm{FE}}^{(b)}=0.5\,$W/Km for a monolayer
thickness of $5$\thinspace\AA \ \cite{Li2015}) because of the simultaneous low
thermal conductivity of the ferroelectric and high available current density
in graphene. This should be easily observable close to the edges, even when
some heat current leaks from the ferroelectric into the graphene. Inversely, a
temperature gradient in the ferroelectric generates a charge current in
graphene, i.e., a nonlocal ferron-drag thermopower. $s_{D}=\pi_{D}%
/T_{0}=1.23\,\mathrm{\mu}$V/K at $T_{0}=298\,$K. However, this number is at
least an order of magnitude smaller than that of a single graphene
\cite{Zuev2009, Wei2009}.

For sufficient thermal isolation between the ferroelectric and graphene layers
the figure of merit of the ferron drag thermoelectric device can be
defined and estimated as
\begin{equation}
\left(  ZT\right)  _{D}=\frac{\sigma s_{D}^{2}T_{0}}{\kappa_{\mathrm{FE}}%
}=2.6\times10^{-3}%
\end{equation}
where $\sigma=1.38\times10^{-3}\,$S is the electrical conductivity with the
nearby ferroelectric at $n_{0}=10^{13}$\thinspace cm$^{-2}$ \cite{SM}. This
$\left(  ZT\right)  _{D}$ is comparable to that of graphene \cite{Reshak2008}
but it may be engineered to become larger by, e.g., optimizing the electron
density of graphene as shown in figure~\ref{Fig-2} or stacking $m$
ferroelectric monolayers with $\left(  ZT\right)  _{D}\propto m$ as long as
all of them stay in the range of the dipolar interaction. The predicted
substantial FOM in spite of the small $s_{D}$ relies on beating the
Wiedemann-Franz Law that hinders conventional thermoelectric devices: The
small heat conductivity in the ferroelectric does not depend on the electric
conductivity in the conductor, which is large in graphene in spite of the
additional ferron scattering \cite{SM}.

According to Supplemental Material B \cite{SM} the current drag is not
specific for ferrons: the expressions are identical for in-plane longitudinal
and out-of-plane polarized polar optical phonons, except for the difference in
the frequency dispersions and other relevant parameters. We therefore
encourage search for thermoelectric effects in any highly polarizable
insulator. The attraction of using ferroelectrics is strong dependence and
control of larger effects by temperature and applied electric field as well as
non-volatile switching of the ferroelectric order. The critical enhancement of
the electrical resistance at the Curie temperature is also unique for ferroelectrics.

\emph{Conclusion:} We predict significant non-local ferron-drag thermoelectric
effects in bilayers of ferroelectric insulators and conductors that are
separated by a small distance $d$. A remote gate-field controlled Peltier
effect can be detected by standard thermography and would prove the existence
of the ferron quasiparticles in ferroelectrics. The results can be readily
extended to the limit $d=0$ corresponding to van der Waals conducting
ferroelectrics, known as ferroelectric metals, in which electric polarization
and mobile electrons coexist \cite{Zhou2020}, and the ferroelectrics with
in-plane spontaneous polarization. In the dipole approximation of the
ferroelectric charge dynamics, the mobile electrons cannot screen the
perpendicular ferroelectric order nor couple to the longitudinal ferrons. We
may expect a strong coupling to the transverse ferrons, however, with
associated interesting thermoelectric phenomena presently under investigation.
Our work opens a new strategy for the design of thermoelectric devices that
are not bound by the Wiedemann-Franz Law.

\emph{Acknowledgements:} We acknowledge the helpful discussions with Ryo
Iguchi. JSPS KAKENHI Grant No. 19H00645 supported P.T. and G.B and Grant No.
22H04965 supported G.B. and K.U. K.U. also acknowledges support by JSPS
KAKENHI Grant No. 20H02609 and JST CREST \textquotedblleft Creation of
Innovative Core Technologies for Nano-enabled Thermal
Management\textquotedblright\ Grant No. JPMJCR17I1.

%%%%%%%%%%%%%%%%%%%%%%%%%%%%

\end{document}

\title{Supplementary Material to \textquotedblleft Nonlocal Drag
Thermoelectricity Generated by a Ferroelectric
Heterostructures\textquotedblright}
\author{Ping Tang$^{1}$}
\author{Ken-ichi Uchida$^{2,3}$}
\author{Gerrit E. W. Bauer$^{1,3,4,5,6}$}
\affiliation{$^1$WPI-AIMR, Tohoku
University, 2-1-1 Katahira, Sendai 980-8577, Japan}
\affiliation{$^2$National Institute for Materials Science, Tsukuba 305-0047, Japan}
\affiliation{$^3$Institute for Materials Research, Tohoku University,
2-1-1 Katahira, Sendai 980-8577, Japan} \affiliation{$^4$Center for
Spintronics Research Network, Tohoku University, Sendai 980-8577, Japan}
\affiliation{$^5$Zernike Institute for Advanced Materials, University of
Groningen, 9747 AG Groningen, Netherlands}
\affiliation{$^6$Kavli Institute for Theoretical Sciences, University of the Chinese Academy of Sciences,
Beijing 10090, China}
\maketitle
%%%%%%%%%%%%%%%%%%%%%%%%%%%%

In this Supplementary Material, we derive the effect of ferron-electron
scatterings on the electrical conductivity in the 2D metal (graphene) (Section
A) and the current drag on normal polar phonons in ferroelectrics or polar
dielectrics (Section B)\textit{.}

\section{A. Ferron fluctuations and electrical conductivity of a nearby metal
film}

The linearized Boltzmann equation within relaxation time
approximation for the distribution function $f_{\mathbf{k}\nu}$ in the metal reads
\begin{equation}
eE_{M}v_{\mathbf{k}\nu}^{(x)}\left(  -\frac{\partial f_{\mathbf{k}\nu}^{(0)}%
}{\partial\varepsilon_{\mathbf{k}\nu}}\right)  =-\frac{f_{\mathbf{k}\nu
}-f_{\mathbf{k}\nu}^{(0)}}{\tau_{e}}+\left(  \frac{\partial f_{\mathbf{k}\nu}%
}{\partial t}\right)  _{\mathrm{int}}\tag{S.A1}\label{Boltz}%
\end{equation}
where $E_{M}$ is the driving field in the $x$-direction, $v_{\mathbf{k}\nu
}^{(x)}=\partial\varepsilon_{\mathbf{k}\nu}/\partial(\hbar k_{x})$ the
corresponding group velocity, $f_{\mathbf{k}\nu}^{(0)}$ the Fermi-Dirac
distribution, $\tau_{e}$ the relaxation time caused by, e.g., impurities,
phonons, etc. within the graphene and the second collision term on the right
hand side stems from the electron-ferron interactions. Fermi's Golden Rule
\begin{align}
\left(  \frac{\partial f_{\mathbf{k}\nu}}{\partial t}\right)  _{\mathrm{int}}=
&  \frac{2\pi}{\hbar}\sum_{\mathbf{q}}\left\vert U_{\mathbf{k}\mathbf{q}}\right\vert ^{2}\left\{
f_{(\mathbf{k}+\mathbf{q})\nu}(1-f_{\mathbf{k}\nu})(N_{\mathbf{q}%
}+1)-f_{\mathbf{k}\nu}(1-f_{(\mathbf{k}+\mathbf{q})\nu})N_{\mathbf{q}%
}\right\}  \delta(\varepsilon_{\mathbf{k}\nu}+\hbar\omega_{q}-\varepsilon
_{(\mathbf{k}+\mathbf{q})\nu})\nonumber\\
+ &  \frac{2\pi}{\hbar}\sum_{\mathbf{q}}\left\vert U_{\mathbf{k}\mathbf{q}}\right\vert
^{2}\left\{  f_{(\mathbf{k}-\mathbf{q})\nu}(1-f_{\mathbf{k}\nu})N_{\mathbf{q}%
}-f_{\mathbf{k}\nu}(1-f_{(\mathbf{k}-\mathbf{q})\nu})(N_{\mathbf{q}%
}+1)\right\}  \delta(\varepsilon_{\mathbf{k}\nu}-\varepsilon_{(\mathbf{k}%
-\mathbf{q})\nu}-\hbar\omega_{q})\tag{S.A2}\label{FGR}%
\end{align}
includes out-scattering (self-energy) and in-scattering (vertex correction)
terms during both the creation and annihilation processes of ferrons. As in the main text, the interband scatterings $(\nu\neq\nu^{\prime})$ have been discarded again. $N_{\mathbf{q}}$ is the ferron distribution in the neighboring ferroelectric
layer. We expand the distribution
\begin{equation}
f_{\mathbf{k}\nu}=f_{\mathbf{k}\nu}^{(0)}+\left(  -\frac{\partial
f_{\mathbf{k}\nu}^{(0)}}{\partial\varepsilon_{\mathbf{k}\nu}}\right)
g_{\mathbf{k}\nu}=f_{\mathbf{k}\nu}^{(0)}+\frac{g_{\mathbf{k}\nu}}{k_{B}%
T}\left(  1-f_{\mathbf{k}\nu}^{(0)}\right)  f_{\mathbf{k}\nu}^{(0)}%
\tag{S.A3}\label{Exp}%
\end{equation}
where $g_{\mathbf{k}\nu}$ represents the nonequilibrium part. In linear
response, we replace the ferron distribution by the equilibrium Planck
distribution $N_{q}^{(0)}$ and substitute Eq.~(\ref{Exp}) into Eq.~(\ref{FGR}%
), i.e.,
\begin{align}
\left(  \frac{\partial f_{\mathbf{k}\nu}}{\partial t}\right)  _{\mathrm{int}}=
&  \frac{2\pi}{\hbar k_{B}T}\sum_{\mathbf{q}}\left\vert U_{\mathbf{k}\mathbf{q}}\right\vert
^{2}(1-f_{(\mathbf{k}+\mathbf{q})\nu}^{(0)})f_{\mathbf{k}\nu}^{(0)}N_{q}%
^{(0)}[g_{(\mathbf{k}+\mathbf{q})\nu}-g_{\mathbf{k}\nu}]\delta(\varepsilon
_{\mathbf{k}\nu}+\hbar\omega_{\mathbf{q}\nu}-\varepsilon_{(\mathbf{k}%
+\mathbf{q})\nu})\nonumber\\
+ &  \frac{2\pi}{\hbar k_{B}T}\sum_{\mathbf{q}}\left\vert U_{\mathbf{k}\mathbf{q}}\right\vert
^{2}(1-f_{\mathbf{k}\nu}^{(0)})f_{(\mathbf{k}-\mathbf{q})\nu}^{(0)}N_{q}%
^{(0)}[g_{(\mathbf{k}-\mathbf{q})\nu}-g_{\mathbf{k}\nu}]\delta(\varepsilon
_{\mathbf{k}\nu}-\varepsilon_{(\mathbf{k}-\mathbf{q})\nu}-\hbar\omega
_{\mathbf{q}\nu}).\tag{S.A4}%
\end{align}
For a uniform charge transport the nonequilibrium part can be written as
$g_{\mathbf{k}\nu}=g_{k\nu}\cos\vartheta_{\mathbf{k}},$ where $\vartheta_{k}$
is the angle between an applied field and wave vector, i.e., a Fermi
distribution shifted in momentum space. As in the main text, we invoke the
quasi-elastic scattering approximation $\delta(\varepsilon_{\mathbf{k}\nu
}-\varepsilon_{(\mathbf{k}\pm\mathbf{q})\nu}\pm\hbar\omega_{q})\approx
\delta(\varepsilon_{\mathbf{k}\nu}-\varepsilon_{(\mathbf{k}\pm\mathbf{q})\nu}= g_{\mathbf{k}\nu}(\cos\theta_{\mathbf{k},\mathbf{k}\pm\mathbf{q}}-1)$ for an isotropic electronic band, and find
\begin{equation}
\left(  \frac{\partial f_{\mathbf{k}\nu}}{\partial t}\right)  _{\mathrm{int}%
}\approx-\frac{f_{\mathbf{k}\nu}-f_{\mathbf{k}\nu}^{(0)}}{\tau_{D}}\tag{S.A5}%
\end{equation}
where
\begin{equation}
\tau_{D}^{-1}=-\frac{2\pi}{\hbar}\sum_{\mathbf{q}}\left\vert U_{\mathbf{k}\mathbf{q}}\right\vert
^{2}2\hbar\omega_{\mathbf{q}}\frac{\partial N_{q}^{(0)}}{\partial\hbar
\omega_{q}}(1-\cos\theta)\delta(\varepsilon_{\mathbf{k}\nu}-\varepsilon
_{(\mathbf{k}+\mathbf{q})\nu})\tag{S.A6}\label{SCF}%
\end{equation}
is the ferron scattering rate and $\theta=\theta_{\mathbf{k},\mathbf{k}%
+\mathbf{q}}$ the angle between $\mathbf{k}$ and $\mathbf{k}+\mathbf{q}$.
Substituting Eq.~(\ref{SCF}) into Eq.~(\ref{Boltz}) leads to
\begin{equation}
f_{\mathbf{k}\nu}=f_{\mathbf{k}\nu}^{(0)}+e\widetilde{\tau}_{e}v_{\mathbf{k}%
\nu}^{(x)}E_{M}\frac{\partial f_{\mathbf{k}\nu}^{(0)}}{\partial\varepsilon
_{\mathbf{k}\nu}}\tag{S.A7}\label{Distr}%
\end{equation}
where $\widetilde{\tau}_{e}=(\tau_{e}^{-1}+\tau_{D}^{-1})^{-1}$ is the
effective transport relaxation time. Substituting $U_{\mathbf{k}\mathbf{q}}$ from the main text, we find for the graphene
\begin{equation}
\tau_{D}^{-1}=-\frac{e^{2}}{16\pi m_{p}\epsilon_{r}^{2}\epsilon_{0}^{2}}\int
d^{2}q\frac{e^{-2dq}}{(1+q_{\mathrm{TF}}/q)^{2}}\frac{\partial N_{q}}%
{\partial\omega_{q}}(1-\cos^{2}\theta)\delta(\varepsilon_{\mathbf{k}%
}-\varepsilon_{\mathbf{k}+\mathbf{q}})\tag{S.A8}\label{Drag}%
\end{equation}
where $q=2k\sin(\theta/2)$.

The exponent limits the momentum transfer to $q<1/\left(  2d\right)  .$ For
the electrons at Fermi level with $q_{\mathrm{TF}}d\gg1$, $k_{F}d\gg1$ and
$d\gg l,$ where $l=\sqrt{g\chi}$ is the ferroelectric coherence length and
$\chi$ the susceptibility,
\begin{equation}
\tau_{D}^{-1}\simeq\frac{3\omega_{0}}{128\pi(k_{F}d)^{2}(q_{\mathrm{TF}}%
d)^{2}(q_{c}d)}\frac{\hbar\omega_{0}}{k_{B}T}\mathrm{csch}^{2}\left(
\frac{\hbar\omega_{0}}{2k_{B}T}\right)  \tag{S.A9}%
\end{equation}
with
\begin{equation}
q_{c}^{-1}=\frac{\chi e^{2}}{v_{F}\epsilon_{r}^{2}\epsilon_{0}^{2}\hbar
}.\tag{S.A10}%
\end{equation}
At the Curie temperature where $\chi\rightarrow\infty$ and $l\rightarrow
\infty$, such that $d\ll l$, we have to reintroduce the ferron dispersion
\begin{equation}
\omega_{q}=\sqrt{\frac{gq^{2}+\chi^{-1}}{m_{p}}}\approx\sqrt{\frac{g}{m_{p}}%
}q.\tag{S.A11}%
\end{equation}
When $k_{B}T\gg(\hbar/d)\sqrt{g/m_{p}}$ the Eq.~(\ref{Drag}) becomes
\begin{equation}
\tau_{D}^{-1}\simeq\frac{1}{32\pi(k_{F}d)(q_{\mathrm{TF}}d)^{2}}\frac{k_{B}%
T}{\varepsilon_{F}}\frac{e^{2}}{\hbar g\epsilon_{r}^{2}\epsilon_{0}^{2}%
}\tag{S.A12}\label{CTc}%
\end{equation}
which is similar to the scattering of electrons by the acoustic phonons in
graphene but with different dependence on the carrier density of graphene
\cite{Hwang2008}. The electrical current density is given by
\begin{equation}
j_{c}=-\frac{4e}{A}\sum_{\mathbf{k}}v_{\mathbf{k}\nu}^{(x)}f_{\mathbf{k}\nu
}\tag{S.A13}%
\end{equation}
where the spin and valley degeneracies have been included. Substituting
Eq.~(\ref{Distr}), we arrive at the series resistor model
\begin{equation}
\frac{1}{\sigma}=\frac{1}{\sigma_{0}}+\frac{1}{\sigma_{D}}\tag{S.A14}%
\end{equation}
where $\sigma_{0}\simeq e^{2}v_{F}^{2}\tau_{e}D_{F}/2$ and $\sigma_{D}\simeq
e^{2}v_{F}^{2}\tau_{D}D_{F}/2$.

\begin{figure}[h]
\centering
\par
\includegraphics[width=16.2cm]{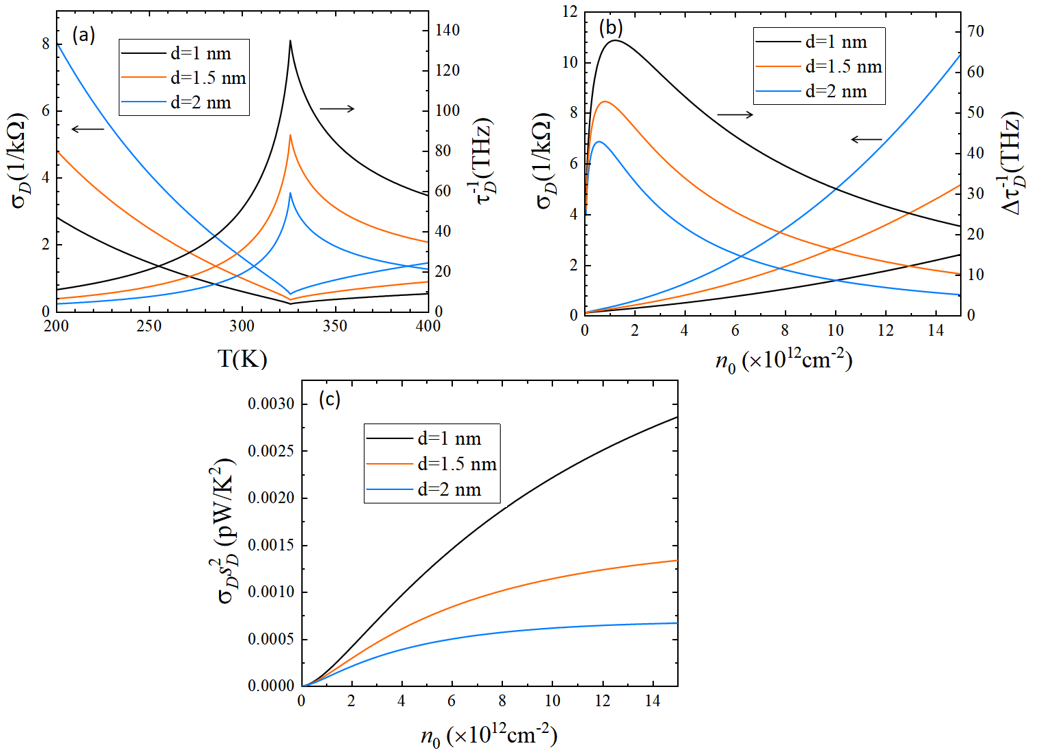}\newline\caption{The scattering rate
$\tau_{D}^{-1}$ at the Fermi level and conductivity $\sigma_{D}$ of graphene
as a function of (a) temperature at a fixed $n_{0}=5\times10^{12}\,$cm$^{-2}$
and (b) carrier density $(n_{0})$ at room temperature for several distances to
a ferroelectric layer with Curie temperature indicated by the vertical dotted
line. (c) The drag power factor $\sigma_{D}s_{D}^{2}$ as a function of $n_{0}$
at room temperature.}%
\label{Fig-SM1}%
\end{figure}

In Fig.~\ref{Fig-SM1} we show a plot of $\tau_{D}^{-1}$ (at the Fermi level)
and $\sigma_{D}$ obtained by computing Eq.~(\ref{Drag}) as a function of (a)
temperature and (b) carrier density $n_{0}$ for the parameters used in the
main text and different interlayer distances. The $\tau_{D}^{-1}$ $(\sigma
_{D})$ is critically enhanced (suppressed) at the Curie temperature $(T_{c})$
determined by Eq.~(\ref{CTc}) and $\sigma_{D}$ increases rapidly with $n_{0}$,
in sharp contrast to the conductivity limited by the acoustic phonon
scattering in graphene \cite{Hwang2008}. While $s_{D}$ varies
non-monotonically with $n_{0}$ (see the main text), the drag power factor
defined as $\sigma_{D}s_{D}^{2}$ and shown in (c) is proportional to $n_{0}$.

At room temperature the acoustic phonon scattering limits $\sigma_{0}$ to $\sim 10^{-2}\,$S in high-quality graphene \cite{Hwang2008}, so $\sigma\approx
\sigma_{D}=1.38\times10^{-3}\,$S and $\sigma s_{D}^{2}\approx2.21\times
10^{-3}\,$pW/K$^{2}$ for $n_{0}=10^{13}\,$cm$^{-2}$ used in the main text.

\section{B. Drag effect by polar optical phonons}

In both ferroelectrics and dielectrics conventional polar optical phonons do
not carry net electric dipoles, but still cause harmonic polarization
fluctuations $\delta\mathbf{P}$ and bound charges $-\boldsymbol{\nabla}%
\cdot\delta\mathbf{P}$ that interact with a proximity metal with electron
density $n(\mathbf{r})$ by
\begin{equation}
\mathcal{H}_{\mathrm{int}}=\frac{e}{4\pi\epsilon_{r}\epsilon_{0}}\int
d\mathbf{r}\int d\mathbf{r}^{\prime}\frac{n(\mathbf{r})\boldsymbol{\nabla
}\cdot\delta\mathbf{P}(\mathbf{r}^{\prime})}{|\mathbf{r}-\mathbf{r}^{\prime}|}
\tag{S.B1}%
\end{equation}
where $e>0$. To be specific we focus on in-plane polarizations ($\delta
\mathbf{P}_{\parallel}$) below, since the treatment of out-of-plane
fluctuations is the same as in the main text. The free energy cost by such
fluctuations can be modeled by
\begin{equation}
\delta F=\frac{1}{2} \int d\mathbf{r}_{\parallel} \left(  g_{\parallel
}(\boldsymbol{\nabla}\delta\mathbf{P}_{\parallel})^{2} +\alpha_{\parallel}
(\delta\mathbf{P}_{\parallel})^{2}-\mathbf{E}_{dip}\cdot\delta\mathbf{P}%
_{\parallel} \right)  \tag{S.B2}%
\end{equation}
where $g_{\parallel}$ ($\alpha_{\parallel}$) accounts for the energy cost by
non-uniform (uniform) fluctuations and $\mathbf{E}_{dip}$ the dipolar electric
field associated with the fluctuations. Note that there is no in-plane
spontaneous polarization except fluctuations. The $\delta\mathbf{P}%
_{\parallel}$ are composed of longitudinal and transverse modes with wave
vector parallel and perpendicular to their fluctuating directions,
respectively. In second quantization,
\begin{equation}
\delta\mathbf{P}_{\parallel}=\sqrt{\frac{\hbar}{2m_{p}^{\parallel}A}}%
\sum_{\mathbf{q},\sigma=L,T}\frac{\mathbf{e}_{\mathbf{q}\sigma}}{\sqrt
{\omega_{\mathbf{q}\sigma}}}a_{\mathbf{q}\sigma}e^{i\mathbf{q}\cdot\mathbf{r}%
}+\mathrm{h.c.} \tag{S.B3}%
\end{equation}
where $m_{p}^{\parallel}$ is the inertia of $\delta\mathbf{P}_{\parallel}$,
$\mathbf{e}_{\mathbf{q}\sigma}$ the polarization of polar phonons with the
creation operator of $a_{\mathbf{q}\sigma}$ and $\sigma=L,T$ represents the
longitudinal ($\mathbf{q}\parallel\mathbf{e}_{\mathbf{q}L}$) and transverse
$(\mathbf{q}\perp\mathbf{e}_{\mathbf{q}T})$ polar optical phonon modes. In
contrast, the ferron modes (out-of-plane fluctuations) that we label
$\omega_{\mathbf{q}}$ are always polarized normal to the plane (wave vector).
For the transverse mode, $\boldsymbol{\nabla}\cdot\delta\mathbf{P}_{\parallel
}=0$, such that $\mathbf{E}_{dip}=0$, and its dispersion relation is thus
given by
\begin{align}
\omega_{\mathbf{q}T}=  &  \sqrt{\frac{\alpha_{\parallel}+g_{\parallel}
\mathbf{q}^{2}}{m_{p}^{\parallel}}}. \tag{S.B4}%
\end{align}
The longitudinal mode carries nonzero $\boldsymbol{\nabla}\cdot\delta
\mathbf{P}_{\parallel}$, which leads to
\begin{equation}
\mathbf{E}_{dip}(\mathbf{r}_{\parallel})=-\frac{1}{4\pi\epsilon_{r}%
\epsilon_{0}}\int d\mathbf{r}_{\parallel}^{\prime}\frac{\boldsymbol{\nabla
}\cdot\delta\mathbf{P}_{\parallel}(\mathbf{r}_{\parallel}^{\prime})}%
{\vert\mathbf{r}_{\parallel}-\mathbf{r}_{\parallel}^{\prime} \vert^{2}}
\tag{S.B5}%
\end{equation}
and the electrostatic energy
\begin{equation}
-\frac{1}{2}\int d\mathbf{r}_{\parallel}\mathbf{E}_{dip}\cdot\delta
\mathbf{P}_{\parallel}=\frac{1}{8\pi\epsilon_{r}\epsilon_{0}}\int
d\mathbf{r}_{\parallel}\int d\mathbf{r}_{\parallel}^{\prime} \frac
{\boldsymbol{\nabla}\cdot\delta\mathbf{P}_{\parallel}(\mathbf{r}_{\parallel
}^{\prime})}{\vert\mathbf{r}_{\parallel}-\mathbf{r}_{\parallel}^{\prime}
\vert^{2}} \delta\mathbf{P}(\mathbf{r}_{\parallel})=\frac{1}{2}\sum
_{\mathbf{q}} \frac{1}{2\epsilon_{r}\epsilon_{0}\vert\mathbf{q}\vert}
\mathbf{q}\cdot\delta\mathbf{P}_{\parallel}(\mathbf{q})\mathbf{q}\cdot
\delta\mathbf{P}_{\parallel}(-\mathbf{q}) \tag{S.B6}%
\end{equation}
where $\delta\mathbf{P}_{\parallel} (\mathbf{q})=\int d\mathbf{r}_{\parallel
}e^{-i\mathbf{q}\cdot\mathbf{r}_{\parallel}}\delta\mathbf{P}_{\parallel
}(\mathbf{r}_{\parallel})$. The corresponding dispersion relation is readily
identified as
\begin{equation}
\omega_{\mathbf{q}L}=\left(  \omega_{\mathbf{q}T}^{2}+\frac{\vert
\mathbf{q}\vert}{2\epsilon_{r}\epsilon_{0} m_{p}^{\parallel}}\right)  ^{1/2}.
\tag{S.B7}%
\end{equation}
Unlike the 3D case, the dipolar interaction lifts the longitudinal mode by a
momentum-dependent term rather than a constant term.

Only the longitudinal ($\mathbf{q}\parallel\mathbf{e}_{\mathbf{q}L}$) polar
phonons couple electrostatically with conduction electrons (with field
operator $c_{\mathbf{k}}$ and $c_{\mathbf{k}}^{\dagger}$), i.e.,
\begin{align}
\mathcal{H}_{\mathrm{int}}  &  =-\sum_{\mathbf{q},\sigma}\frac{ie}%
{2\epsilon_{r}\epsilon_{0}}\sqrt{\frac{\hbar}{2m_{p}^{\parallel}A}}%
\frac{\mathbf{q}\cdot\mathbf{e}_{\mathbf{q}\sigma}e^{-dq}}{q\sqrt
{\omega_{\mathbf{q}\sigma}}}n(\mathbf{-q})a_{\mathbf{q}\sigma}+h.c.\nonumber\\
&  =-\sum_{\mathbf{k},\mathbf{q}}\frac{ie}{2\epsilon_{r}\epsilon_{0}}%
\sqrt{\frac{\hbar}{2m_{p}^{\parallel}A}}\frac{e^{-dq}}{\sqrt{\omega
_{\mathbf{q}L}}}n(-\mathbf{q})a_{\mathbf{q}L}+h.c.. \tag{S.B8}%
\end{align}
The interaction matrix elements are similar to that of the ferrons in the main
text. The drag Peltier coefficient of the polar phonons follows from Eq.~(22)
as
\begin{gather}
\pi_{D}\approx\frac{e\tau_{f}g_{\parallel}\hbar^{3}D_{F}^{2}}{32\delta_{\nu
}(m_{p}^{\parallel}) ^{2}\epsilon_{r}^{2}\epsilon_{0}^{2}n_{0}k_{F}k_{B}T}%
\int_{0}^{2k_{F}}\frac{q^{2}dq}{\sqrt{1-(q/2k_{F})^{2}}}\frac{\cos^{2}%
(\theta/2)e^{-2dq}}{(1+q_{\mathrm{TF}}/q)^{2}}\mathrm{csch}^{2}\left(
\frac{\hbar\omega_{qL}}{2k_{B}T}\right) \nonumber\\
\sim\mathrm{csch}^{2}\frac{\omega_{0L}}{2k_{B}T}\sim\frac{1}{\omega_{0L}^{2}%
},\,\,\,(\hbar\omega_{0L}\ll k_{B}T). \label{SB4}\tag{S.B9}%
\end{gather}
Since $\omega_{0L}$ ($> 10\,$THz) is usually much larger than the gap of the
soft ferron mode ($\sim2.9\,$THz for our studied ferroelectric at room
temperature), its relative effect on the drag Peltier coefficient in
ferroelectrics is small. The normal phonons also do not contribute to the
critical enhancement of the transport coefficients at the phase transition.
Moreover, the harmonic polar phonons do not carry electric dipoles, so
$\omega_{0L}$ does not experience a Stark shift by an external electric field.

Eq.~(20) in the main text remains valid for a polar dielectric/graphene
bilayer without ferrons, but $\omega_{q}$ is then the frequency dispersion of
the out-of-plane optical phonons, and should be added to Eq. (\ref{SB4}) for the
total effect.